\documentclass[aps,twocolumn,prb,amsmath,amssymb,showpacs,superscriptaddress]{revtex4}
\pdfoutput=1

\usepackage{graphicx}
\usepackage{bm}
\usepackage[breaklinks=true,colorlinks,citecolor=blue,linkcolor=blue,urlcolor=blue]{hyperref}
\usepackage{ulem}
\usepackage{color}
\newcommand{\us}{\uparrow}
\newcommand{\ds}{\downarrow}

\begin{document}

\title{Local density of states in metal -- topological superconductor hybrid systems}

\author{Marco Gibertini}
\email{marco.gibertini@sns.it} 
\affiliation{NEST, Scuola Normale Superiore and Istituto Nanoscienze-CNR, I-56126 Pisa, Italy}
\author{Fabio Taddei}
\affiliation{NEST, Istituto Nanoscienze-CNR and Scuola Normale Superiore, I-56126 Pisa, Italy}
\author{Marco Polini}
\affiliation{NEST, Istituto Nanoscienze-CNR and Scuola Normale Superiore, I-56126 Pisa, Italy}
\author{Rosario Fazio}
\homepage{http://qti.sns.it/}
\affiliation{NEST, Scuola Normale Superiore and Istituto Nanoscienze-CNR, I-56126 Pisa, Italy}

\date{\today}

\begin{abstract}
We study by means of the recursive Green's function technique the local density-of-states of (finite and semi-infinite) multi-band spin-orbit 
coupled semiconducting nanowires in proximity to an $s$-wave superconductor and attached to normal-metal electrodes. 
When the nanowire is coupled to a normal electrode, the zero-energy peak, 
corresponding to the Majorana state in the topological phase, broadens with increasing transmission between the 
wire and the leads, eventually disappearing for ideal interfaces. 
Interestingly, for a finite transmission a peak is present also in the normal electrode, even though it has a smaller 
amplitude and broadens more rapidly with the strength of the coupling. Unpaired Majorana states can survive close to a topological phase transition even when the number of open channels (defined in the absence of superconductivity) is even. We finally study the Andreev-bound-state spectrum in superconductor-normal metal-superconductor junctions and find that in multi-band nanowires the distinction between topologically trivial and non-trivial systems based on the number of zero-energy crossings is preserved.
\end{abstract}
\pacs{74.78.Na,74.45.+c,71.10.Pm,74.78.-w}

\maketitle

\section{Introduction}

Since the first prediction~\cite{Majorana1937} of real solutions to the Dirac equation, known as Majorana fermions, 
there have been many attempts to demonstrate their occurrence in nature~\cite{WilczekNatPhys2009}, but a 
clear evidence is still lacking. Besides the natural search for these elusive particles in high-energy physics, it has been 
recently suggested that Majorana fermions can exist as exotic excitations in certain condensed-matter systems~\cite{BeenakkerArxiv2011,AliceaArxiv2012}. 
Such solid-state realizations include fractional quantum Hall states at filling factor $\nu=5/2$~\cite{MooreNuclPhys1991}, 
$p$-wave superconductors and superfluids~\cite{ReadPRB2000,IvanovPRL2001}, three-dimensional topological insulators in 
proximity to $s$-wave superconductors~\cite{FuPRL2008}, as well as spin-orbit coupled semiconductors in a 
magnetic field with proximity-induced $s$-wave superconducting pairing~\cite{Kitaev2001,SauPRL2010,
AliceaPRB2010,SauPRB2010,OregPRL2010,LutchynPRL2010}. The importance of finding Majorana fermions 
in condensed-matter systems is not only related to their fundamental interests.  It is also rooted in the 
non-Abelian braiding statistics of these particles which could be exploited as a basis for decoherence-free 
topological quantum computation~\cite{NayakRMP2008}. 

In this paper we focus on a specific proposal  realized with spin-orbit-coupled semiconducting nanowires in 
proximity to an $s$-wave superconductor (S) and subjected to an in-plane magnetic field~\cite{SauPRB2010,LutchynPRL2010,OregPRL2010} 
(a system that, for the sake of simplicity, will be henceforth termed ``S-nanowire"). 
This wire can support Majorana-fermion bound states at its ends when 
parameters such as chemical potential, magnetic field, and superconducting pairing are properly tuned~\cite{SatoPRB2010}. 
The S-nanowire is said to be in the topological phase when Majorana bound states are present, while it is topologically trivial 
otherwise. 

In order to assess the presence of Majorana fermions in such solid-state systems, it is of primary importance to predict 
clear signatures of the topological phase, which could be then used to guide experiments. A very relevant quantity 
is the Local Density-of-States (LDOS), which can be accessed in scanning tunneling microscopy. 
LDOS calculations have already been carried out in Refs.~\onlinecite{PotterPRB2011} and~\onlinecite{StanescuArxiv2011}, but 
were restricted to finite-size S-nanowires in which superconductivity is uniform along the wire. The influence of normal segments inside a one-dimensional S-nanowire was investigated in Ref.~\onlinecite{SauArxiv2011}.
In this work we exploit the recursive Green's function technique to study the LDOS of both finite and semi-infinite multi-band S-nanowires in the presence of normal electrodes (NS junctions), as well as Supercoductor-Normal metal-Superconductor (SNS) junctions. In this manner we have been able to i) reproduce previous results~\cite{PotterPRB2011,StanescuArxiv2011,SauArxiv2011}, ii) generalize them to account for the coupling to normal electrodes and/or for a finite width of the wire, and iii) consider more complicated structures like SNS junctions. 

The paper is organized as follows. In Sec.~\ref{sec:two} we describe the model and we introduce the corresponding 
Hamiltonian. The phase diagram is presented and new analytical expressions for the phase boundaries are reported. 
Section~\ref{sec:three} is devoted to the discussion of our numerical results for the LDOS. We start in Sec.~\ref{subsec:isolated} 
with the case of a finite-size multi-band S-nanowire~\cite{PotterPRB2011}. We then consider a single NS junction between 
semi-infinite leads in Sec.~\ref{subsec:NS}. This situation can arise for instance when the nanowire is
only partially in proximity to a bulk superconductor so that part of
the nanowire is in the normal state. Here we also study the evolution of the LDOS across the topologically trivial/topologically non-trivial phase transition. In Sec.~\ref{subsec:SNS} we turn our attention to a SNS junction, illustrating numerical results for the LDOS and for the Andreev-bound-state spectrum. Finally, in Section~\ref{sec:conclusion} we draw our main conclusions.
 
\section{Model Hamiltonian}
\label{sec:two}

Our calculation of the LDOS is based on a recursive Green's function method especially designed for tight-binding Hamiltonians. In 
order to apply this method to the present case, we describe the semiconducting nanowire as a square lattice with a finite width $W$ 
in the direction perpendicular to the axis of the wire (the ${\hat {\bm y}}$-direction) and lattice constant $a$.
Assuming the presence of Rashba-type spin-orbit (SO) coupling, of strength $\alpha$, and a Zeeman field $V$ along the wire 
($\hat {\bm x}$-direction), the lattice discretization of the usual continuum-model Hamiltonian~\cite{LutchynPRL2010,OregPRL2010,AkhmerovPRL2011} 
reads~\cite{SatoPRB2010}
\begin{eqnarray}\label{eq:hamilt}
\hat {\cal H}_{\rm RZ} &=& \sum_{i,j} \left[h_{\rm RZ}(i,j)\right]_{\sigma \sigma'} \hat c^\dag_{i,\sigma}\hat c_{j,\sigma} \nonumber\\
&=& -t\sum_{\langle i,j\rangle,\sigma}\hat c^\dag_{i,\sigma}\hat c_{j,\sigma} + (\varepsilon_0-\mu)\sum_{ i,\sigma}\hat c^\dag_{i,\sigma}\hat c_{i,\sigma} \nonumber\\
&& + i\alpha \sum_{\langle i,j\rangle,\sigma,\sigma'} (\nu'_{ij}\sigma^x_{\sigma\sigma'} - \nu_{ij}\sigma^y_{\sigma\sigma'})\hat c^\dag_{i,\sigma}\hat c_{j,\sigma'} \nonumber\\
&&  + V \sum_{i,\sigma,\sigma'}\sigma^x_{\sigma\sigma'}\hat c^\dag_{i,\sigma}\hat c_{i,\sigma'}~.
\end{eqnarray}
Here $\varepsilon_0= 4 t$ is a uniform on-site energy which sets the zero of energy, $\sigma^{i}$ are spin-1/2 Pauli matrices, $\nu_{i j} = \hat{\bm x} \cdot \hat {\bm d}_{i j}$, and $\nu'_{i j} = \hat{\bm y} \cdot \hat {\bm d}_{i j}$, with $\hat {\bm d}_{i j} = ({\bm r}_i - {\bm r}_j)/|{\bm r}_i - {\bm r}_j|$ being the unit vector connecting site $j$ to site $i$.

If we now allow sections of the nanowire to be in contact with a bulk $s$-wave superconductor, the proximity effect induces a non-vanishing superconducting pairing in these sections so that the complete Hamiltonian becomes
\begin{equation}\label{eq:FULLhamilt}
\hat {\cal H} = \hat {\cal H}_{\rm RZ} + \hat{\cal H}_{\rm S}~,
\end{equation}
where
\begin{equation}\label{eq:SChamilt}
\hat {\cal H}_{\rm S} = \sum_{ i}\left[\Delta(i)~\hat c^\dag_{i,\us}\hat c^\dag_{i,\ds} + \text{H.c.}\right]~.
\end{equation}
For simplicity we assume $\Delta(i)$ to be piecewise constant, with $|\Delta(i)|=\Delta$ in the regions in contact with the superconductor and $\Delta(i)=0$ otherwise.
Moreover, we assume that a barrier is present at the boundary between proximized and non-proximized sections, leading to a decrease of the value of the hopping energy $t$ and of the SO coupling $\alpha$ by the same factor $\gamma$.~\cite{note:gamma}

Finally, it is convenient to introduce the Nambu spinors $\hat \Psi_i = (\hat c_{i,\us},\hat c_{i,\ds},\hat c^\dag_{i,\ds},-\hat c^\dag_{i,\us})^T$ and re-write the Hamiltonian~\eqref{eq:FULLhamilt} in the form
\begin{equation}
\hat {\cal H} = \frac{1}{2}\sum_{i,j}\hat \Psi^\dag_i~{\cal H}_{\rm BdG}(i,j)~\hat\Psi_{j}~,
\end{equation}
where 
\begin{equation}\label{eq:BdGhamilt}
{\cal H}_{\rm BdG}(i,j) =
\begin{pmatrix}
h_{\rm RZ}(i,j) & \Delta(i)~\delta_{ij} \\
\Delta(i)~\delta_{ij} &  - \sigma^y~h_{\rm RZ}^*(i,j)~\sigma^y
\end{pmatrix}
\end{equation}
is the Bogoliubov-de Gennes (BdG) Hamiltonian~\cite{deGennes}.

The tight-binding Hamiltonian \eqref{eq:BdGhamilt} will be the starting point of our analysis.

\subsection{Phase diagram}
%
\begin{figure}
\includegraphics[width=0.8\linewidth]{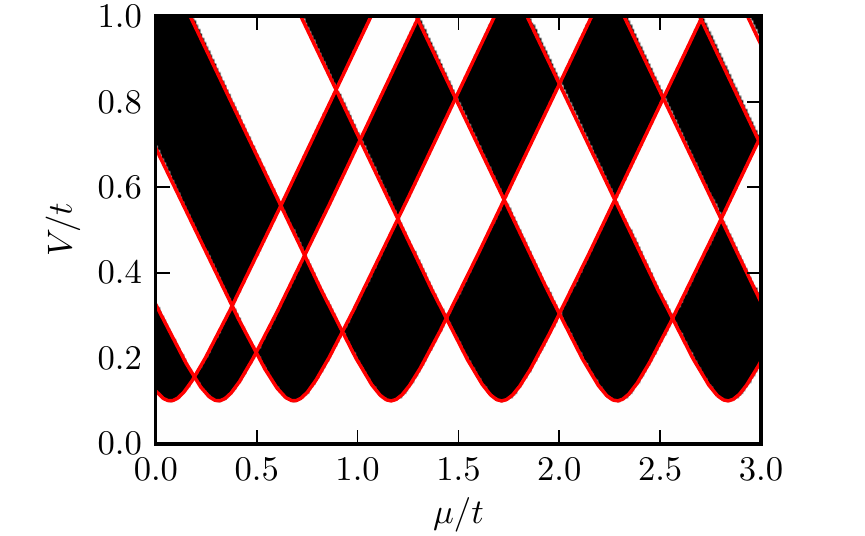}
\caption{(Color online) Phase diagram of an infinite superconducting wire as a function of the chemical potential $\mu$ and the Zeeman field $V$. White regions correspond to a trivial system (${\cal Q} = +1$), while dark regions identify the topologically non-trivial phase (${\cal Q} = -1$). Red thick lines illustrate the prediction in Eq.~\eqref{eq:phasebound} for the phase boundaries. These results refer to the following set of parameters: $W/a=10$, $\alpha/t=0.1$, and  $\Delta/t=0.1$.\label{fig:PhDiag}}
\end{figure}

Before proceeding with the study of the LDOS we need to know in which regions of parameter space we should expect Majorana fermions. This problem has been addressed in Refs.~\onlinecite{LutchynPRL2010} and~\onlinecite{OregPRL2010} in the strictly one-dimensional case and in the continuum limit ($a\to 0$): the S-nanowire is in the topological phase when $|V|>\sqrt{\mu^2 + \Delta^2}$ while it is in the trivial phase otherwise. Thus the phase boundary occurs along the line implicitly defined by
\begin{equation}\label{eq:phasebound1D}
V^2 = \mu^2 + \Delta^2~.
\end{equation}
The phase diagram of multiband ($W/a\neq1$) nanowires has been investigated numerically in Refs.~\onlinecite{PotterPRB2011} and \onlinecite{AkhmerovPRL2011}. The phase of a uniform system can be determined for instance from the evaluation of the following Pfaffian formula~\cite{GhoshPRB2010} for the topological invariant 
\begin{equation}\label{eq:topinvariant}
{\cal Q} = {\rm sign}\left\{{\rm Pf}[{\cal H}_{\rm BdG}(0)\sigma^y\tau^y]~{\rm Pf}[{\cal H}_{\rm BdG}(\pi/a)\sigma^y\tau^y]\right\}~,
\end{equation}
where ${\cal H}_{\rm BdG}(k_x)$ is the Fourier transform of the BdG Hamiltonian in Eq.~\eqref{eq:BdGhamilt}, while $\tau^i$ are Pauli matrices acting on the particle-hole degrees-of-freedom. In Fig.~\ref{fig:PhDiag} we report numerical results for a given system ($W/a=10$, $\alpha/t=0.1$, and $\Delta/t=0.1$) as a function of the chemical potential $\mu$ and Zeeman field $V$, obtained using an algorithm developed by Wimmer~\cite{WimmerArxiv2011}. The topologically trivial phase corresponds to ${\cal Q} = +1$ (white regions), while the non-trivial one is signaled by ${\cal Q} = -1$ (dark regions).
The phase boundaries in this figure (red thick lines) have been derived analytically and are given by the following result:
\begin{equation}\label{eq:phasebound}
(\mu-\varepsilon_0-\varepsilon_\lambda \pm 2t)^2 + \Delta^2 = V^2~.
\end{equation}
Here $\varepsilon_0+\varepsilon_\lambda \mp 2t$ are the eigenenergies of ${\cal H}_{\rm BdG}(k)$ for $k=0, \pi/a$, respectively, when $V = \Delta = \mu = 0$. The following expression holds for the energies $\varepsilon_\lambda$:
\begin{equation}\label{eq:energies}
\varepsilon_\lambda = -2 \sqrt{t^2+\alpha^2}\cos\left(\frac{\lambda\pi}{n+1}\right)
\quad \lambda=1,\dots,n=W/a ~.
\end{equation}
A thorough derivation of Eqs.~\eqref{eq:phasebound} and \eqref{eq:energies} is given in Appendix~\ref{app:A}.

\section{LDOS - Numerical Results}\label{sec:three}

\begin{figure}
\includegraphics[width=0.8\linewidth]{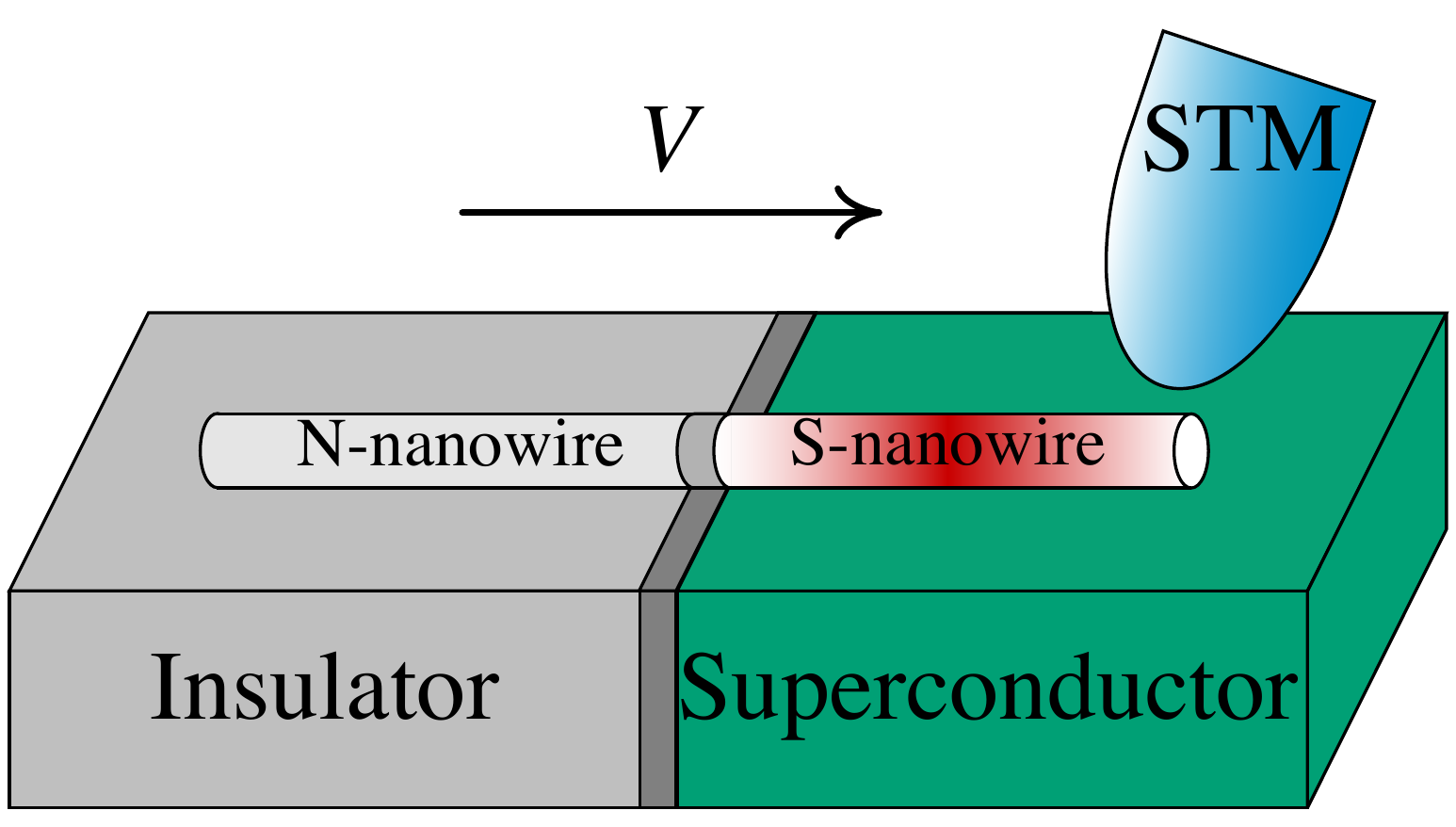}
\caption{Experimental setup adopted in measurements of the LDOS of a nanowire (in this case comprising a normal electrode and a superconducting segment separated by a barrier). The differential conductance obtained from the current between the tip of the scanning tunneling microscope (STM) and the nanowire is proportional to the LDOS of the nanowire according to Eq.~\eqref{eq:LDOS-STM}.\label{fig:setup}}
\end{figure}

As we mentioned in the Introduction, the LDOS can be accessed in experiments using scanning tunneling microscopy (STM). The STM experimental setup  is sketched in Fig.~\ref{fig:setup}. When the metallic tip of the STM is moved close to the nanowire a tunneling current can flow. By locally measuring this current $I$ as a function of the tip-sample bias voltage $V$, the LDOS at a given position ${\bm r}$ and energy $E$ can be reconstructed from the differential conductance at $eV=E$~\cite{TersoffPRB1985,Chen1993}
\begin{equation}\label{eq:LDOS-STM}
 \frac{d I}{d V} ({\bm r}, eV) \propto{\cal N}({\bm r},E=eV)~.
\end{equation}
As a consequence, it is particularly interesting to investigate the LDOS theoretically and make predictions that can be tested in experiments.
%

In what follows the LDOS is computed through the standard relation
\begin{equation}
{\cal N}({\bm r},E) = -\frac{1}{2 \pi}\Im m~\{{\rm Tr}[G({\bm r},E)]\}~,
\end{equation}
where $G({\bm r},E)$ is the Green's function and the factor 2 in the denominator is introduced to avoid a double counting of particle and hole degrees-of-freedom intrinsic in the BdG formalism. We have computed $G({\bm r},E)$ using a recursive Green's function technique similar to the one adopted in Ref.~\onlinecite{PotterPRB2011}, suitably generalized to include the effects of semi-infinite leads~\cite{SanvitoPRB1999}. For simplicity, in the following we fix the width $W=10~a$, the SO coupling strength $\alpha=0.1~t$ and the superconducting pairing $\Delta=0.1~t$.

\subsection{Isolated S-nanowire}\label{subsec:isolated}

\begin{figure}[t]
\includegraphics[width=1.0\linewidth]{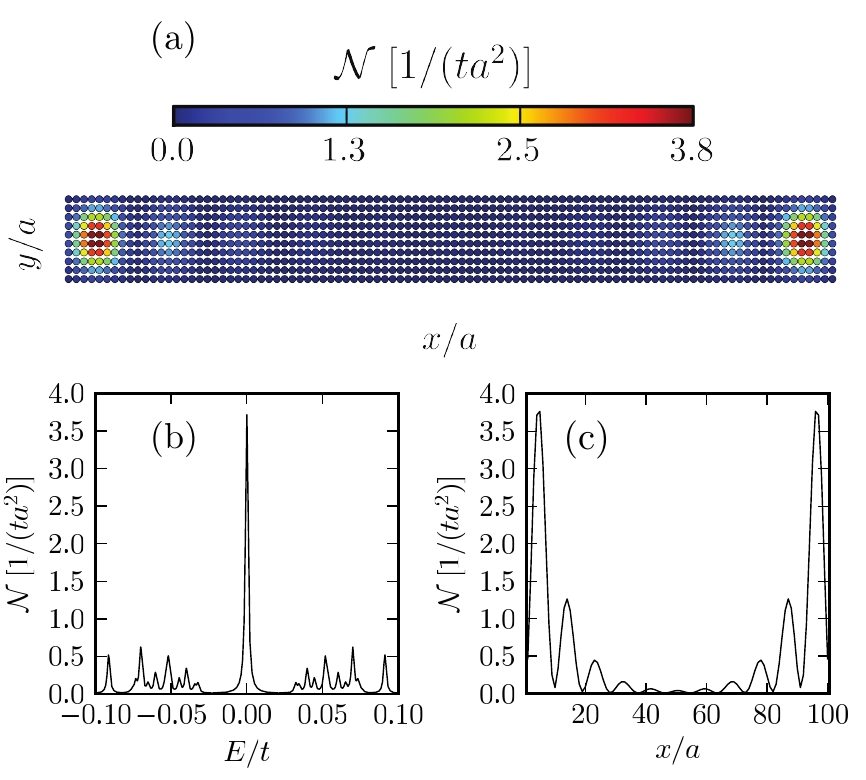}
\caption{(Color online) (a) LDOS of an isolated superconducting nanowire in the topologically non-trivial phase ($\mu/t = 0$, $V/t=0.2$) at an energy very lose to the chemical potential ($E\simeq 0$). Bound states at both ends of the wire are apparent. (b) LDOS at a given position ($x/a=4$, $y/a=5$) as a function of energy. A sharp peak corresponding to a Majorana bound state is present at $E=0$. (c) LDOS at $E\simeq 0$ as a function of $x$ along the middle of the wire ($y/a =5$). \label{fig:fsize-nontriv}}
\end{figure}

Let us first consider an isolated S-nanowire of finite length ($L=100~a$). This situation has been already addressed before~\cite{PotterPRB2011,StanescuArxiv2011} and it is considered here for the sake of reference.
We consider two cases: i) $\mu=0$ and $V/t=0.2$ (with one open channel in the absence of superconducting pairing) and ii) $\mu=0$ and $V/t = 0.6$ (with two open channels).
According to Fig.~\ref{fig:PhDiag}, in case i) the wire is topologically non-trivial, while in case ii) the wire is topologically trivial.
For case i), Fig.~\ref{fig:fsize-nontriv}(a) shows that the LDOS at an energy close to the chemical potential is characterized 
by the presence of bound states at both ends of the wire.
The presence of Majorana bound states appears as  a sharp peak at zero energy in the LDOS as a function of energy 
at a given position in space [see Fig.~\ref{fig:fsize-nontriv}(b)].
According to Fig.~\ref{fig:fsize-nontriv}(c), these Majorana bound states have oscillating wavefunctions which decay 
exponentially inside the bulk of the S-nanowire with a typical length scale (effective superconducting coherence length) $\xi\approx 10~a$.
%
%
%
\begin{figure}[t]
\includegraphics[width=1.0\linewidth]{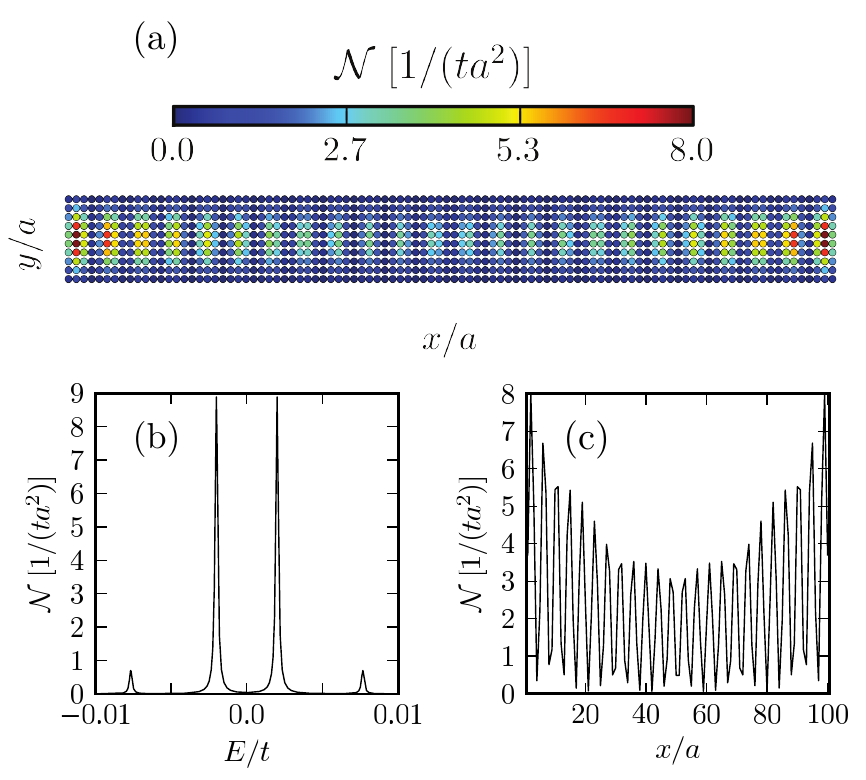}
\caption{(Color online) (a) LDOS of an isolated superconducting nanowire in the topologically trivial phase ($\mu/t = 0$, $V/t=0.6$) at $E/t\simeq 0.002$. (b) LDOS at a given position ($x/a=2$, $y/a=5$) as a function of energy. Owing to the coupling between the two Majorana end states, two Dirac-fermion modes appear at finite energy ($E/t\simeq 0.002$). (c) LDOS at $E/t\simeq 0.002$ as a function of $x$ along the middle of the wire ($y/a =5$). \label{fig:fsize-triv}}
\end{figure}
%

On the contrary, in case ii) where the wire is topologically trivial (and presents two transverse channels in the absence of superconducting pairing) the LDOS at the chemical-potential energy is almost zero throughout the wire, while it shows spatial features only at finite energies [see Fig.~\ref{fig:fsize-triv}(a) for $E/t\simeq\pm 0.002$, where $E$ is measured from the chemical potential].
In particular, as shown in Fig.~\ref{fig:fsize-triv}(c), the LDOS oscillates and decreases moving toward the center of the wire, the length scale of the exponential drop,  $\xi\approx 30~a$, being much larger than for case i).
Now, if the two channels were decoupled, two Majorana modes would have appeared at each end of the wire, one for each open channel.
However, since the two transverse channels are actually coupled in the wire, a single fermion at each end appears at a finite energy~\cite{PotterPRB2011}, as though coming from the hybridization of the two Majorana modes.
Fig.~\ref{fig:fsize-triv}(b) shows one pair of peaks at $E/t\simeq\pm 0.002$ and another pair (with smaller amplitude) at $E/t\simeq\pm 0.007$.
The presence of two pairs of peaks is due to the long coherence length which allows the two localized fermions at the ends of the wire to strongly hybridize, lifting the parity degeneracy of the system. We have checked that for a longer wire ($L/a=400$) the overlap between the fermions vanishes and the two peaks at $E/t\simeq 0.002$ and $E/t\simeq 0.007$ merge into a single double-degenerate peak at energy $E/t\simeq 0.001$.
The latter energy depends on the width of the nanowire~\cite{KellsArxiv2011}.

\subsection{S-nanowire attached to a normal electrode}\label{subsec:NS}

In this Section we analyze the impact of a normal lead attached to the S-nanowire.This situation can arise for instance  when the nanowire is only partially in proximity to a bulk superconductor so that part of the nanowire is in the normal state (as shown in Fig.~\ref{fig:setup}).
In order to get rid of finite-size effects, we consider a semi-infinite S-nanowire coupled to a normal lead. The case of a finite-length S-nanowire coupled to two normal leads at both ends does not yield additional significant information. 

%
\begin{figure}
\includegraphics[width=1.0\linewidth]{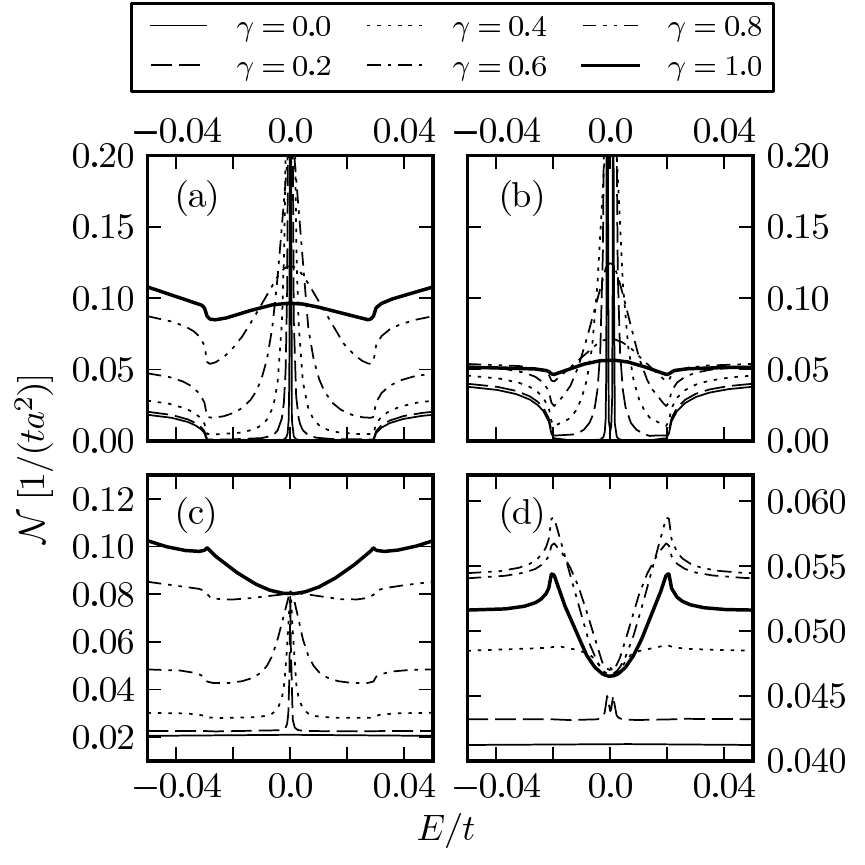}
\caption{LDOS of a superconducting nanowire coupled to a normal lead as a function of energy, at fixed positions in space, in the non-trivial [$\mu/t = 0$ and $V/t=0.2$, panel (a) and (c)],  and trivial [$\mu/t = 0$, $V/t=0.6$, panel (b) and (d)] phase. Panels (a) and (b) refer to a position just inside the superconducting part ($x/a=51$), while panels (c) and (d) refer to a position just inside the normal part of the junction ($x/a=50$). The interface is at $x/a=50.5$.\label{fig:vstranspNS-nontriv}}
\end{figure}
\begin{figure}
\includegraphics[width=1.0\linewidth]{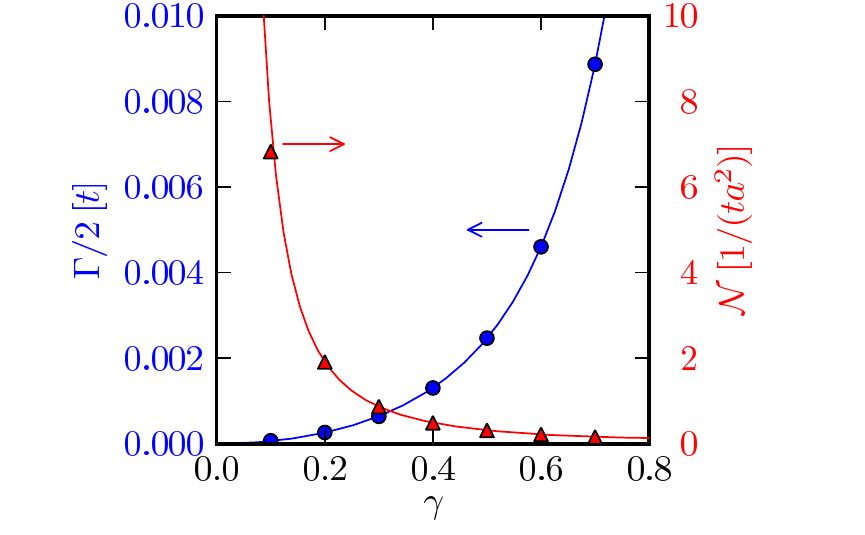}
\caption{(Color online) Fitted values [through Eq.~(\ref{eq:lorentz})] of the half width at half maximum $\Gamma/2$ (blue circles) and height at zero energy ${\cal N}$ (red triangles) of the peak in the LDOS at a position just inside the superconducting nanowire in the non-trivial phase. Solid lines are just guides for the eye.
\label{fig:fit}}
\end{figure}

In Fig.~\ref{fig:vstranspNS-nontriv} we plot the LDOS, at different positions, as a function of energy for several values of the barrier strength $\gamma$. The two plots on the left [(a) and (c)] are for a non-trivial nanowire with $\mu/t = 0$ and $V/t=0.2$, while the two plots on the right [(b) and (d)] refer to a topologically trivial nanowire with $\mu/t = 0$ and $V/t=0.6$.
Moreover, the top panels [(a) and (b)] refer to a position close to the interface in the S-nanowire, while the bottom panels [(c) and (d)] to a position close to the interface in the normal lead.
When $\gamma$ is small the LDOS in the S-nanowire presents a finite gap $E_{\rm g}$ which is just a fraction of the superconducting pairing $\Delta$ owing to the presence of the Zeeman field. Namely, $E_{\rm g}\simeq 0.03~t=0.3~\Delta$ for the non-trivial nanowire and $E_{\rm g} \simeq 0.018~t=0.18~\Delta$ for the trivial case.
Within the gap, Fig.~\ref{fig:vstranspNS-nontriv}(a) shows a sharp Majorana peak at zero energy which broadens as $\gamma$ increases~\cite{SenguptaPRB2011,HutasoitPRB2011}, and eventually disappears when $\gamma\to1$.
Such a peak can be fitted with the following Lorentzian function
\begin{equation}\label{eq:lorentz}
{\cal N}({\bm r},E)\simeq {\cal N} \frac{\left(\Gamma/2\right)^2}{E^2 + \left(\Gamma/2\right)^2}~,
\end{equation}
where $\Gamma/2$ is the half width at half maximum and ${\cal N}$ is the height at zero energy.
The result of the fit is reported in Fig.~\ref{fig:fit}: $\Gamma/2$ and ${\cal N}$ are plotted as functions of $\gamma$.
Remarkably, $\Gamma/2$ depends only very weakly on the position in the S-nanowire where the LDOS is calculated.

Interestingly, the Majorana peak is present also in the LDOS of the normal lead [Fig.~\ref{fig:vstranspNS-nontriv}(c)] as long as $\gamma$ is 
not exactly zero: the peak is still clearly distinguishable up to $\gamma\simeq 0.6$.
Moreover, singularities develop at energies corresponding to $\pm E_{\rm g}$ as $\gamma$ tends to 1.

In the trivial phase, the LDOS in the S-nanowire at a position close to the interface [Fig.~\ref{fig:vstranspNS-nontriv}(b)] presents a single pair of peaks at $\gamma=0$ (as compared to Fig.~\ref{fig:fsize-triv}), since, being semi-infinite, the ends of the S-nanowire are sufficiently far away to be decoupled.
Such peaks quickly broaden as $\gamma$ increases and eventually merge into a single peak at $\gamma\simeq0.3$ (a further increase of $\gamma$ leads to the disappearing of the peak).
As a result, as long as the coupling between the S-nanowire and the normal lead is not too strong, non-trivial and trivial phases can be distinguished from a measurement of the LDOS in the S-nanowire given a sufficiently large energy resolution (in the present case higher than 1\% of the pairing $\Delta$).
On the contrary, only a very weak double-peak structure is visible in the LDOS in the normal lead at a position close to the interface and for small values of $\gamma$ [see Figs.~\ref{fig:vstranspNS-nontriv}(d)].

We mention that important clues on the topological phase of a S-nanowire coupled to a normal electrode can be also retrieved using a different transport setup with respect to the one depicted in Fig.~\ref{fig:setup}.
Indeed, instead of considering the current from the STM tip to the sample, it would in principle be possible to study directly transport through the NS junction present in the nanowire. Even though a theoretical analysis of this configuration is beyond the scope of the present work, in Appendix~\ref{app:B} we report on the low-bias conductance of the NS junction. We find that, in the tunneling limit, the approximate quantization of the conductance can be exploited to identify the topological phase of the nanowire~\cite{SenguptaPRB2011,LawPRL2009,FlensbergPRB2010}.  We also remark that, even for transparent barriers, clear signatures of the presence of Majorana fermions can be extracted from the transport properties of the NS junction provided that a quantum point contact is present close to the interface~\cite{WimmerNJP2011}.

%

\begin{figure}
\includegraphics[width=1.0\linewidth]{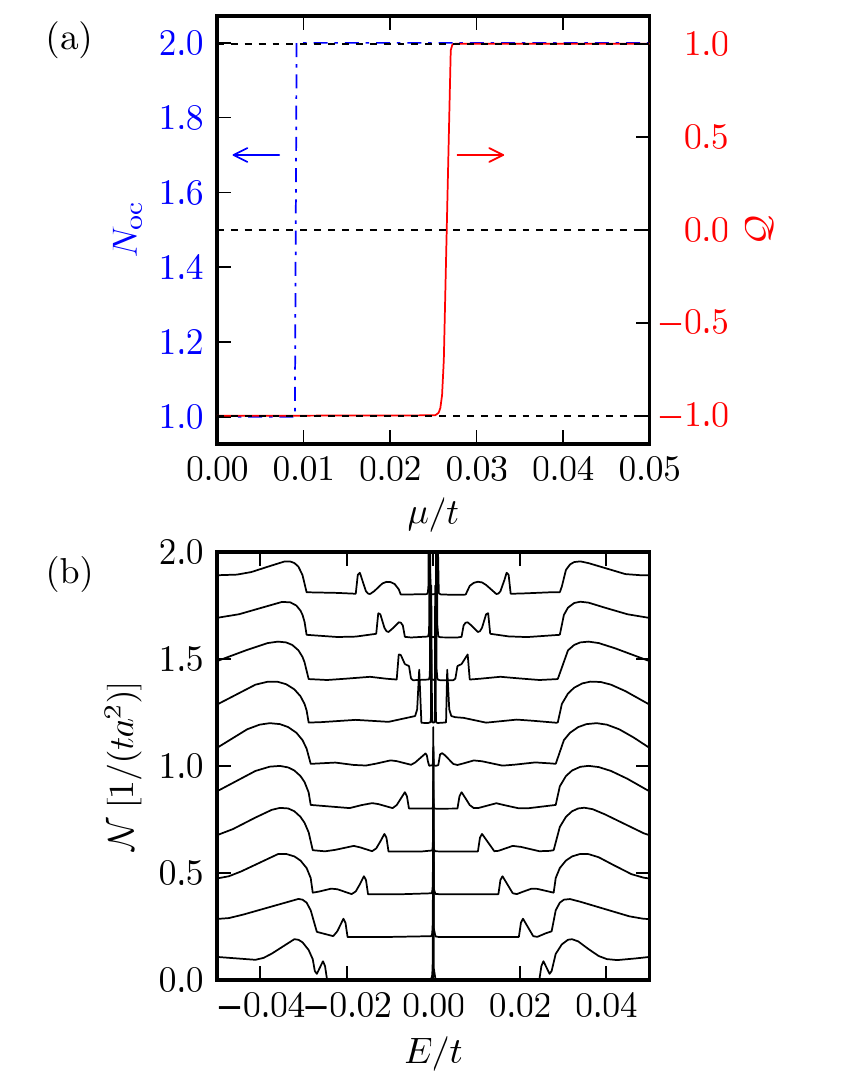}
\caption{(Color online) (a) Topological invariant ${\cal Q}$ (red solid line) and number of open transverse channels $N_{\rm oc}$ (blue dash-dotted line) as functions of the chemical potential $\mu$ for a system with $V/t=0.3$. A phase transition occurs at $\mu/t\simeq 0.026$. (b) LDOS at a position inside a semi-infinite superconducting nanowire for different values of the chemical potential $\mu$ keeping $V/t=0.3$ fixed. Results have been offset vertically for clarity, with an increasing value of the chemical potential from bottom ($\mu/t=0$) to top ($\mu/t=0.05$).  \label{fig:phasetrans}}
\end{figure}

Let us now analyze how the LDOS changes when the S-nanowire is driven through a topological phase transition~\cite{YamakageArxiv2011}.
In Fig.~\ref{fig:phasetrans}(a) we plot the topological invariant ${\cal Q}$ [calculated using Eq.~(\ref{eq:topinvariant})] and the number of open channels ($N_{\rm{oc}}$) as functions of the chemical potential $\mu$ for a fixed Zeeman field $V=0.3~t$.
The S-nanowire goes through a transition, from the non-trivial (${\cal Q}=-1$) to the trivial (${\cal Q}=+1$) phase, at $\mu\simeq 0.026~t$.
Interestingly, the number of channels increases from 1 to 2 at a much smaller value of the chemical potential $\mu\simeq 0.01~t$ such that the non-trivial phase persists even in the presence of 2 open channels. This observation is in apparent contradiction with the intuitive picture adopted in the literature to explain the phase diagram of superconducting nanowires, {\it i.e.} that a pair of Majorana fermions at the ends of the wire is associated with each open channel and that pairs of Majorana fermions on the same end can couple and form complex (Dirac) fermions. Accordingly, there should be a single isolated Majorana fermion at each end of the nanowire whenever the number of open channels is odd. On the other hand, this intuitive picture should be treated with care, simply because one concept (presence of Majorana fermions) is related to a superconducting wire while the other (number of open channels) to a normal one. Indeed, already in Ref.~\onlinecite{PotterPRB2011} it was noticed that the system can be in the topologically trivial phase even when $N_{\rm oc}$ is odd. Here, we are observing the complementary situation in which the topological phase persists when $N_{\rm oc}$ is even. We believe that the underlying explanation is the same for both cases: the failure of the intuitive picture reported above to explain the whole phase diagram. As a consequence, we remark that the topological invariant is not necessarily in a one-to-one correspondence with the parity of the number of open channels.
In Fig.~\ref{fig:phasetrans}(b) the LDOS as a function of energy is shown at a position inside the S-nanowire ($x=50~a$, measured from the interface, and $y=W/2=5~a$)  when $\gamma$ is close to zero. Different curves correspond to different values of the chemical potential, with a vertical offset proportional to $\mu$.
The Majorana peak splits into two Dirac-fermion peaks at finite energy as the chemical potential moves through the phase transition.
Besides, we also observe that the effective gap $E_{\rm g}$ initially decreases for increasing $\mu$, then vanishes at the phase transition, and thereafter increases again. 
Similar results (concerning  the differential tunneling conductance at one end of a nanowire) have been recently presented in Ref.~\onlinecite{StanescuArxiv2011}.

%
%

\subsection{SNS structure}\label{subsec:SNS}

\begin{figure}
\includegraphics[width=1.0\linewidth]{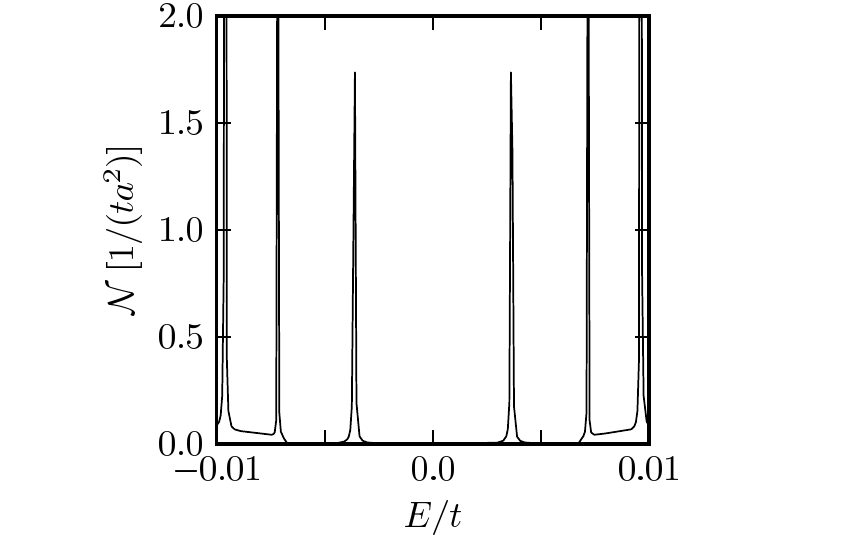}
\caption{LDOS as a function of energy at a given position close to the interface inside the left superconducting nanowire. Sharp peaks corresponding to Andreev bound states appear both within the effective gap $E_{\rm g}$ $(\simeq 0.007~t)$ and inside the continuum.  These results refer to the following set of parameters: $\Delta\varphi=2.0$, $\mu/t =0$, $V/t = 0.8$, and $L/a=0$. \label{fig:ldosAndreev}}
\end{figure}

Let us now consider two semi-infinite S-nanowires connected through a normal nanowire (N-nanowire).
For simplicity we assume transparent barriers at the interfaces and we set 
\begin{equation}
\Delta(i)\equiv\Delta(x/a) = 
\begin{cases}
\Delta~e^{i\varphi_{\rm L}} &x<-L/2\\
0 &|x|\leq L/2\\
\Delta~e^{i \varphi_{\rm R}} &x>L/2\\
\end{cases}
~,
\end{equation}
{\it i.e.} we allow for a finite phase difference $\Delta\varphi = \varphi_{\rm R}-\varphi_{\rm L}$ between the superconducting paring in the right and left S-nanowires, $L$ being the length of the N-nanowire.
Independently of the topological phase of the system, Andreev bound states (ABSs) appear as sharp peaks in the LDOS (see  Fig.~\ref{fig:ldosAndreev}) and the ABS spectrum can thus be reconstructed from the energies at which such peaks occur.
As we shall discuss at length below, what differs between topologically trivial and non-trivial phases is the parity of the number of zero-energy crossings in the ABS spectrum, which is related to the presence or absence of such a crossing at $\Delta\varphi=\pi$ (protected by fermion parity).
This result is in agreement with similar calculations performed for a strictly one-dimensional system~\cite{LutchynPRL2010,TewariArxiv2011}, for a two-dimensional SNS junction~\cite{Black-SchafferArxiv2011}, and for a quantum spin-Hall insulator sandwiched between superconducting leads~\cite{FuPRB2009}. 
%

\begin{figure}
\includegraphics[width=1.0\linewidth]{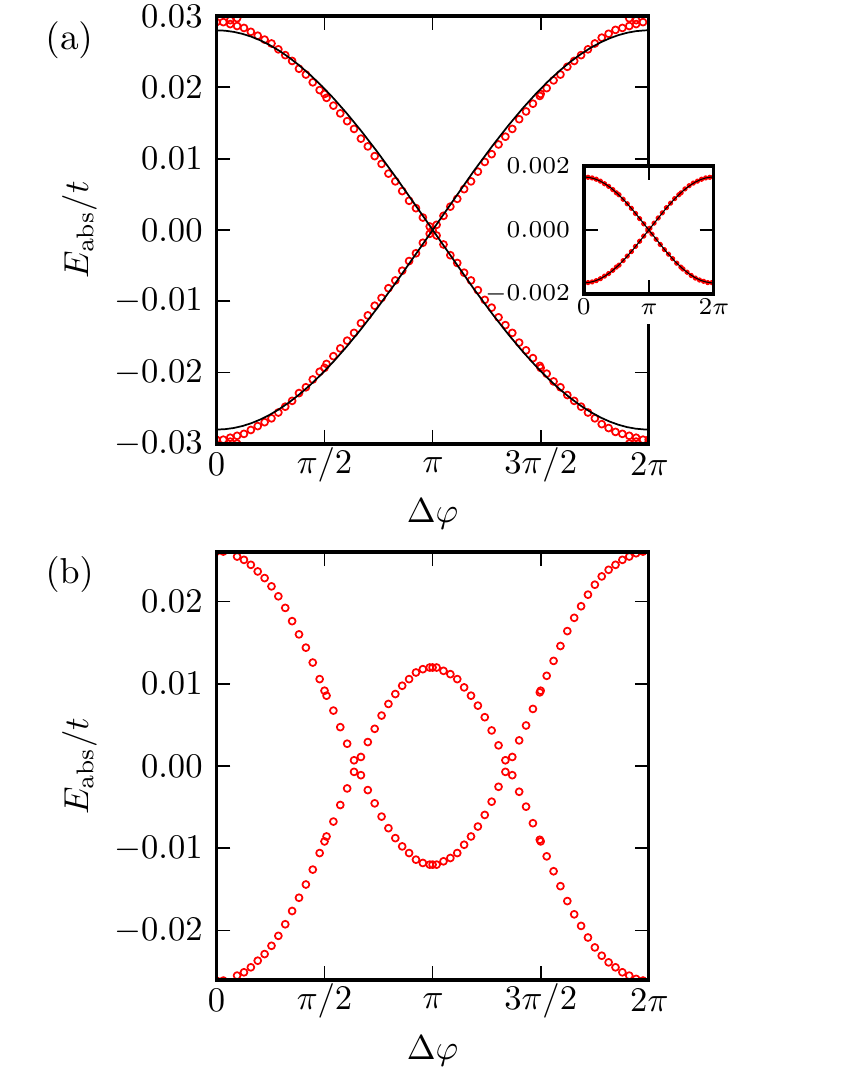}
\caption{(a) Andreev bound state energies (red empty circles) as a function of the phase difference $\Delta\varphi$ for topologically non-trivial superconducting nanowires ($\mu/t=0$, $V/t=0.2$, and $\gamma=1$). The solid line is a best-fit using the expression in Eq.~\eqref{eq:nontrivialABS}.Inset: same as in the main panel but for a non transparent barrier between the superconducting nanowires ($\gamma=0.1$). (b) same as in the top panel but for topologically trivial superconducting nanowires ($\mu/t=0.15$ and $V/t=0.1$).  Both panels refer to a SNS junction with $L/a=0$.
\label{fig:snsSJ}}
\end{figure}

We start by considering the short-junction limit, $L\ll \xi$ (where $\xi\propto t a /E_{\rm g}$ is the effective superconducting coherence length), in which we have a single ABS for each open channel.
$L=0$ is a particular case in which the N-nanowire is absent and we have a step-like jump from $\varphi_{\rm L}$ to $\varphi_{\rm R}$ in the phase of a S-nanowire.
In Fig.~\ref{fig:snsSJ} we show, for $L=0$, the ABS energies as a function of $\Delta\varphi$ for a topologically non-trivial [panel (a) for $\mu/t=0$ and $V/t=0.2$] and trivial [panel (b) for $\mu/t=0.15$ and $V/t=0.1$] S-nanowires.
In both cases we have a single ABS (with its opposite energy counterpart).
The main difference between the two cases is that the number of zero-energy crossings is odd in the non-trivial phase while it is even in the trivial case, in agreement with Ref.~\onlinecite{LutchynPRL2010}.
Furthermore, even though the overall periodicity of the ABS spectrum is $2\pi$ in both cases, the periodicity of a single branch of the ABS spectrum is $4\pi$ only for the non-trivial situation.
In particular, the change of $\Delta\varphi$ by $2\pi$ at a fixed energy leads to the swapping of an ABS with its charge-conjugate state which has opposite fermion parity. In terms of Josephson current this leads to the fractional Josephson effect~\cite{Kitaev2001} and can be interpreted either as a $4\pi$-periodicity or as a two-valuedness of the Josephson current. For simplicity in the following we will address the topologically non-trivial ABS spectrum as $4\pi$-periodic. 

It is now interesting to compare the plot in Fig.~\ref{fig:snsSJ}(a) with the ABS spectrum relative to a short, one-dimensional, SNS Josephson junction with $p_x + i p_y$ superconducting order parameter.
An expression for the latter has been obtained, under the Andreev approximation (which assumes an order parameter much smaller than the Fermi energy), in Ref.~\onlinecite{KwonLTP2004}:
\begin{equation}\label{eq:nontrivialABS}
E_{\rm abs} = \pm \Delta_{\rm abs}~\sqrt{{\cal T}}~\cos\left(\Delta\varphi/2\right)~,
\end{equation}
where ${\cal T}$ is the transmission probability of the N region and $\Delta_{\rm abs}$ is an effective order parameter for the ABS.
The solid line in Fig.~\ref{fig:snsSJ}(a) shows the result of a best fit, with respect to $\Delta_{\rm abs}$, once ${\cal T}$ is set to 1.
The agreement between the numerical results (red empty circles) and the fit is quite good, despite the fact that the S-nanowire behaves effectively as a $(p_x + i p_y)$-wave superconductor only in the limit of large Zeeman fields, with respect to $\Delta$ (in the present case $V=0.2~t$ and $\Delta=0.1~t$)~\cite{note:pwave}. In the inset we show that, contrary to a non-topological SNS junction~\cite{BeenakkerPRL1991}, the zero-energy crossing survives even when the barrier between the superconductors is not transparent ($\gamma=0.1$). In accordance with Eq.~\eqref{eq:nontrivialABS}, the only effect of a transparency ${\cal T}<1$ is to reduce the band-width of the ABS spectrum preserving the crossing at $\Delta\varphi=\pi$. Indeed, with respect to the non-topological case discussed for instance in Ref.~\onlinecite{BeenakkerPRL1991}, here the ABSs have different fermion parity, {\it i.e.} they correspond to states with an even or odd number of fermions, respectively. Since fermion parity is a conserved quantity in this case, the two ABSs can not couple (giving rise to the opening of a gap) even when the transparency of the system is lowered.

\begin{figure}
\includegraphics[width=1.0\linewidth]{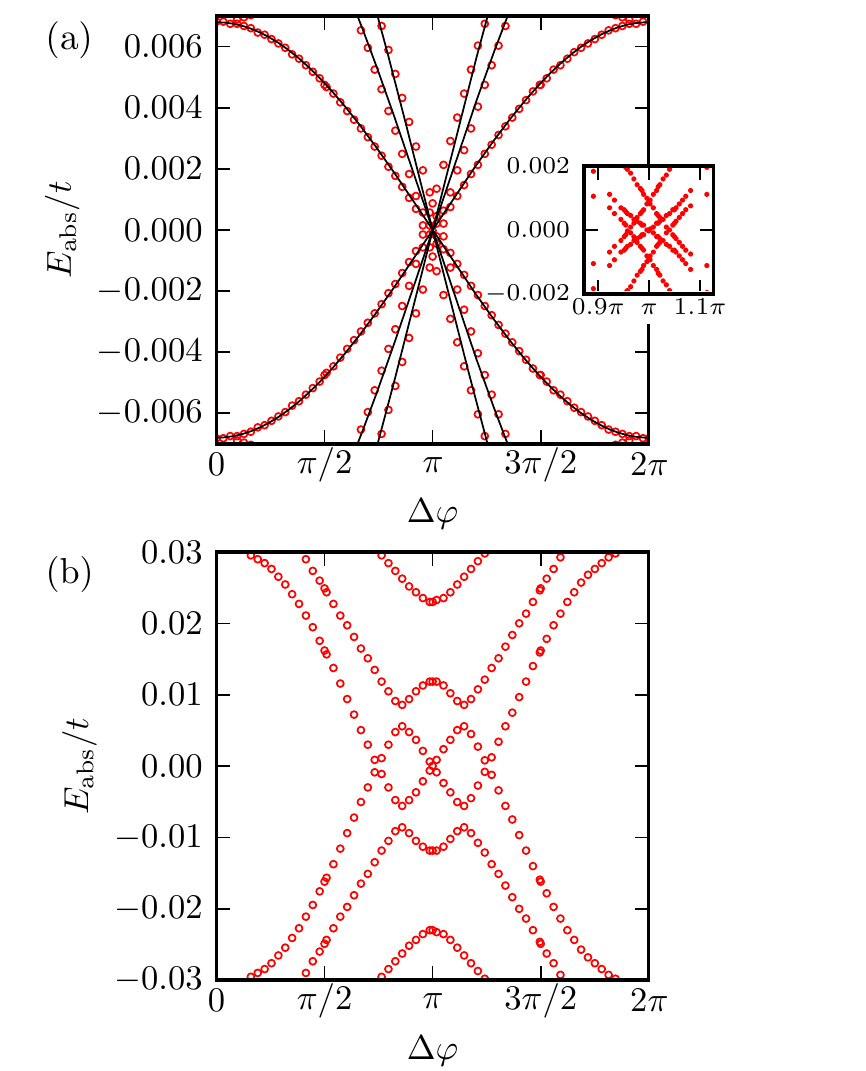}
\caption{(a) Andreev bound state energies (red empty circles) as a function of the phase difference $\Delta\varphi$ for topologically non-trivial superconducting nanowires when $\mu/t=0$ and $V/t=0.8$. The solid lines are fits to the function in Eq.~\eqref{eq:nontrivialABS} with ${\cal T}=1$ for three different values of $\Delta_{\rm abs}$ for the three Andreev bound states. The inset shows a magnification of the region close to $\Delta\varphi=\pi$ at small energies. (b) same as in the panel a but for $\mu/t=1.7$ and $V/t=0.2$. Both panels refer to a SNS junction with $L/a=0$.
\label{fig:sns-nontrivial}} 
\end{figure}

Let us now address the case of larger Zeeman fields, where, however, the situation is complicated by the fact that by increasing $V$ also the number of open channels increases.
In Fig.~\ref{fig:sns-nontrivial}(a) the ABS spectrum, as a function of $\Delta\varphi$, is shown for a topologically non-trivial S-nanowire with $\mu=0$ and $V=0.8~t$, where three transverse open channels are allowed in the absence of superconducting pairing.
The spectrum presents three ABS branches and an odd (three) number of crossings at zero energy, as expected for the non-trivial case. 
We fit separately the corresponding three ABS branches with Eq.~\eqref{eq:nontrivialABS} allowing three different values of $\Delta_{\rm abs}$, to be interpreted as effective gaps for each open channel, one for each branch.
As shown in the plot, the fits (solid lines) apparently approximate quite well the numerical data (red empty circles).
The agreement, however, is completely lost for energies close to zero [see the inset of Fig.~\ref{fig:sns-nontrivial}(a)]: i) only one ABS branch crosses zero energy at $\Delta\varphi=\pi$, ii) the spectrum shows complicated avoided crossings which shift the zero-energy crossing of the two remaining ABS branches to $\Delta\varphi\neq\pi$.
As already discussed in Sec.~\ref{subsec:NS}, the Majorana modes, which would localize at a given end if the three open channels were not coupled, strongly hybridize at $\Delta\varphi=\pi$ giving rise to a true Majorana mode plus a pair of Dirac excitations at finite energy.
Moreover, we notice that the periodicity of all three ABS branches is $4\pi$ and the peaks in the LDOS connected to the two ABSs with larger energy appear (see also Fig.~\ref{fig:ldosAndreev}) even above the bulk gap $E_{\rm abs}$ ($\simeq 0.007~t$ in this case).

The situation changes completely in the case of a smaller Zeeman field ($V/t=0.2$), still in the topologically non-trivial phase, while keeping the number of open channels equal to three ({\it i.e.} by setting $\mu/t =1.7$).
The corresponding spectrum, reported in Fig.~\ref{fig:sns-nontrivial}(b), presents again an odd (three) number of zero-energy crossings, but shows gap-like features: no fitting with Eq.~\eqref{eq:nontrivialABS} is now possible.
This is due to the fact that now the effective $p_x+ip_y$ description no longer holds since ABSs with both chiralities are present~\cite{AliceaPRB2010}, so that an inter-band $s$-wave pairing amplitude can mix them.
In this case, one of the ABS branches has periodicity $4\pi$ while the remaining two have periodicity $2\pi$. We still expect a fractional Josephson effect. 
%

\begin{figure}
\includegraphics[width=1.0\linewidth]{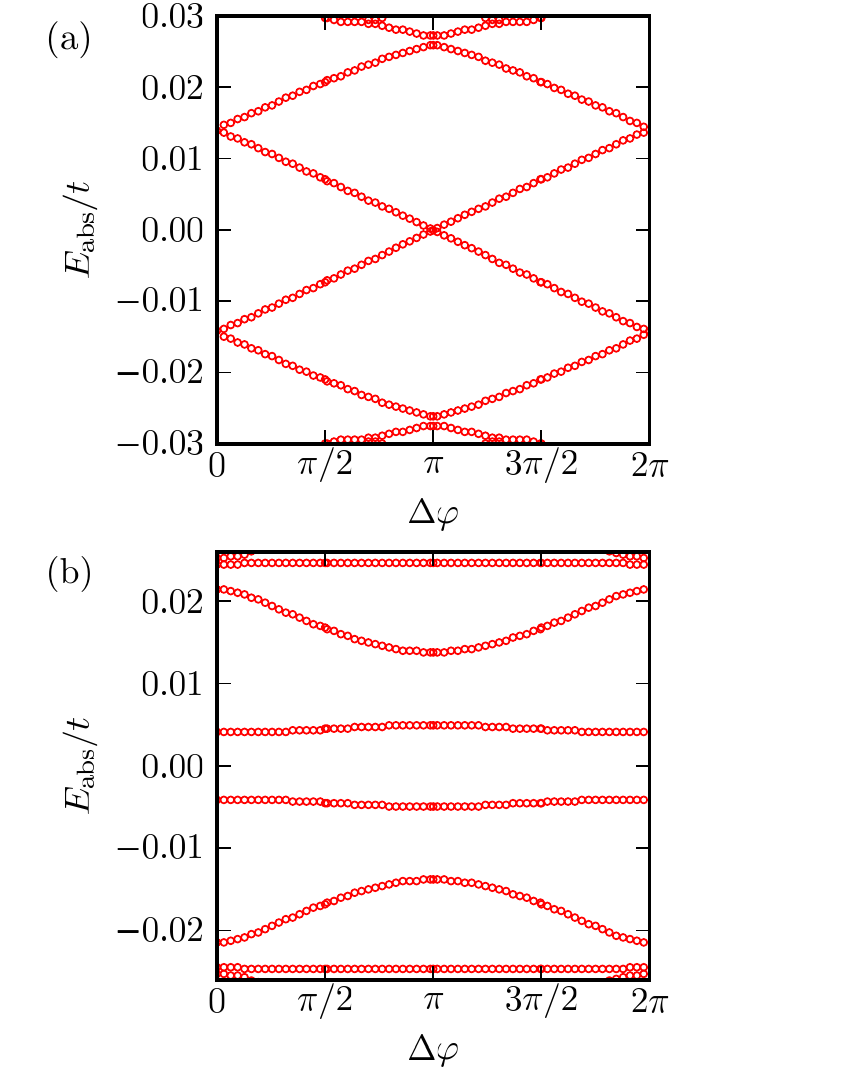}
\caption{(a) Andreev bound state energies (red empty circles) as a function of the phase difference $\Delta\varphi$ for topologically non-trivial superconducting nanowires ($V/t=0.2$ and $\mu/t=0$). (b) same as in the top panel but for topologically trivial superconducting nanowires ($V/t=0.05$ and $\mu/t=0$).  Both panels refer to the long-junction limit ($L/a=50$).
\label{fig:snsLJ}} 
\end{figure}

If we increase the length of the N-nanowire toward the long-junction limit ($L\gg\xi$), the number of ABSs increases but the topological phase of the S-nanowire can still be detected from the number of zero-energy crossings in the interval $\Delta\varphi\in [0,2\pi]$~\cite{LutchynPRL2010,FuPRB2009}.
This is shown in Fig.~\ref{fig:snsLJ} where the ABS spectrum is plotted for $L/a=50$ (long-junction limit) when the S-nanowires are in the non-trivial phase [panel (a) for $V/t=0.2$ and $\mu/t=0$] and in the trivial phase [panel (b) for $V/t=0.05$ and $\mu/t=0$].
As in the short-junction limit, the periodicity of at least one branch of the ABS spectrum is $4\pi$ only in the non-trivial case.
%
%

\section{Conclusions}\label{sec:conclusion}

In this paper we have considered a multi-band semiconducting nanowire subjected to spin-orbit coupling, superconducting pairing and a longitudinal Zeeman field.
Depending on the values of such parameters, the nanowire presents a non-trivial topological phase in which a pair of Majorana modes, at an energy equal to the chemical potential, are localized at its ends.
We have first derived an analytic expression for the phase boundaries of an infinitely long multi-band nanowire.
We have then numerically calculated and analyzed the local density-of-states of such nanowires in the case when they are coupled to normal regions (such as electrodes or links) and we have compared the topologically non-trivial and trivial phases in different situations.
When the nanowire is coupled to a normal electrode we have found that the peak in the LDOS at zero energy (with respect to the chemical potential), corresponding to the Majorana mode, broadens with increasing coupling strength to the electrode, eventually disappearing for a transparent interface.
Interestingly, for finite coupling the peak is also present in the normal electrode, though being of smaller amplitude and broadening more rapidly with the strength of the coupling.
In the trivial phase, and when the nanowire possesses two open channels in the absence of superconducting pairing, a pair of peaks at finite energies appears as due to the hybridization of the two Majorana modes that would exist if the two channels of the nanowire were not coupled.
Such peaks broaden with increasing coupling strength to the normal electrode, eventually merging for sufficiently large coupling.
In the normal electrode only weak features survive.
From the analysis of the topological phase transition, driven by varying the chemical potential at fixed Zeeman field, we have found that the nanowire remains in the topologically non-trivial phase even after the number of open channels goes from one to two. This suggests that, contrary to the intuitive picture often referred to in the literature, the one-to-one correspondence between the topological invariant and the parity of the number of open channels is only approximate and should be treated with care.
We have then considered the situation in which two semi-infinite nanowires (kept at different superconducting phases) are connected through a normal link of length $L$.
Independently of the topological phase the density-of-states presents peaks due to Andreev bound states whose position in energy depends on the superconducting phase difference $\Delta\varphi$.
While in the trivial phase the number of zero-energy crossings is even, in the topologically non-trivial phase this number is odd owing to the presence of a fermion-parity-protected crossing at $\Delta\varphi=\pi$. 
This difference in the parity of the number of zero-energy crossings reflects the presence of at least one branch of Andreev-bound-state energy which is $4\pi$-periodic (instead of the usual $2\pi$-periodicity), leading to the so-called fractional Josephson effect.
This anomalous $4\pi$-periodicity of the Josephson current has been usually introduced in strictly one-dimensional systems while we have checked that it survives also in a multi-band nanowire, in agreement with Ref.~\onlinecite{LawPRB2011}. 

\acknowledgments
We would like to acknowledge fruitful discussions with C.W.J. Beenakker and D. Rainis. 
This work has been supported by the EU FP7 Programme under Grant Agreement  
No. 234970-NANOCTM, No. 248629-SOLID, No. 233992-QNEMS, No. 238345-GEOMDISS, and No. 215368-SEMISPINNET.

\appendix

\section{Derivation of the phase diagram}\label{app:A}

We first need to determine the eigenvalues and eigenvectors of ${\cal H}_{\rm BdG}(k_x)$ for $k_x=0$ and  $\pi/a$, when $V = \Delta = 0$. The difference between $k_x=0, \pi/a$ is just a shift $\pm 2 t$ in the chemical potential, {\it i.e.}
\begin{equation}
{\cal H}_{\rm BdG}(k_x=0,\pi/a) = {\cal H}_0 + [\varepsilon_0-(\mu\pm 2 t)]~\tau_z~.
\end{equation}
Moreover, we are assuming $\Delta=0$, so that 
\begin{equation}
{\cal H}_0 = 
\begin{pmatrix}
{\cal H}_{\rm p}  &  0 \\
0 & -\sigma^y  {\cal H}^*_{\rm p} \sigma^y
\end{pmatrix}~.
\end{equation}
Thus, it suffices to consider the particle Hamiltonian ${\cal H}_{\rm p}$.
For a wire of width $W/a=n$ the characteristic polynomial $L_n(\varepsilon;\alpha)$ of ${\cal H}_{\rm p}$ can be defined recursively as
\begin{equation}\label{eq:recursive}
L_n(\varepsilon;\alpha) = \varepsilon~L_{n-1}(\varepsilon;\alpha) - (t^2+\alpha^2)~L_{n-2}(\varepsilon;\alpha)~,
\end{equation}
with $L_0(\varepsilon;\alpha)=1$ and $L_1(\varepsilon;\alpha)=\varepsilon$. A formal solution to the recursive relation~\eqref{eq:recursive} is given by
\begin{equation}
L_n(\varepsilon;\alpha) = \frac{r_{+}^{n+1} - r_{-}^{n+1}}{\sqrt{\varepsilon^2 - 4 (t^2+\alpha^2)}}
\end{equation}
with
\begin{equation}
r_{\pm} = \frac{\varepsilon \pm \sqrt{\varepsilon^2 - 4 (t^2+\alpha^2)}}{2}~.
\end{equation}
One can easily solve $L_n(\varepsilon;\alpha) = 0$ and find the following expression for the energies $\varepsilon_i$
\begin{equation}
\varepsilon_\lambda = -2 \sqrt{t^2+\alpha^2}\cos\left(\frac{\lambda\pi}{n+1}\right)
\quad \lambda=1,\dots,n~.
\end{equation}
For completeness, we mention that each eigenvalue is doubly degenerate and the corresponding eigenstates are characterized by an amplitude $\psi^{\pm}_{m\sigma}$ on site $m$ and spin $\sigma$ given by
\begin{equation}\label{eq:eigenstates}
\psi^{\pm}_{m,\uparrow} = \frac{1}{\sqrt{n+1}}~e^{\mp i m \theta}~\sin\left(m~\frac{\lambda\pi}{n+1}\right) = \pm \psi^{\pm}_{m,\downarrow}
\end{equation}
where the upper and lower signs refer to the two eigenstates and $\tan\theta=\alpha/t$. Moreover, we observe that even allowing for a finite Zeeman field $V\neq0$, the eigenstates do not change since we are assuming the Zeeman field and the spin-orbit coupling in the transverse direction to be both proportional to $\sigma^x$. On the other hand, if $V\neq0$ the two eigenstates in Eq.~\eqref{eq:eigenstates} are no longer degenerate, but split to energies $\varepsilon_i\pm V$.\\
We now turn to the evaluation of the topological invariant ${\cal Q}$ in Eq.~\eqref{eq:topinvariant} when $\Delta\neq0$. We need to compute the Pfaffians:
\begin{equation}
{\rm Pf}[{\cal H}_{\rm BdG}(k_x)\sigma^y\tau^y]
\quad\text{with}\quad k_x = 0,\pi/a~.
\end{equation}
We first introduce the $2n\times2n$ unitary matrix $U$ of the eigenvectors~\eqref{eq:eigenstates} of ${\cal H}_{\rm p}$ and the associated $4n\times4n$ unitary matrix $U_{\rm BdG}$,
\begin{equation}
U_{\rm BdG} = 
\begin{pmatrix}
U  &  0 \\ 
0  & i\sigma^y~ U^*
\end{pmatrix}~,
\end{equation}
which diagonalizes ${\cal H}_{\rm BdG}(0,\pi/a)$ when $\Delta=0$. Even if when $\Delta\neq0$ the following matrix
\begin{eqnarray}
{\cal D}(k_x) &=&  U_{\rm BdG}^{\dag} {\cal H}_{\rm BdG}(k_x) U_{\rm BdG}
\end{eqnarray}
is not diagonal, it is still useful to introduce it. Indeed, we can then write
\begin{equation}
{\cal H}_{\rm BdG}(k_x)\sigma^y\tau^y = 
 U_{\rm BdG} {\cal D}(k_x) (\sigma^y\tau^yU_{\rm BdG})^{\dag}
\end{equation}
and, since by particle-hole symmetry we have that 
\begin{equation} 
\sigma^y\tau^y U_{\rm BdG} = U_{\rm BdG}^* \tau^x~,
\end{equation}
the Pfaffian can be written as 
\begin{eqnarray}
{\rm Pf}[{\cal H}_{\rm BdG}(k_x)\sigma^y\tau^y] &=& {\rm Pf}[U_{\rm BdG} {\cal D}(k_x) \tau^x U_{\rm BdG}^{T}] \nonumber\\
&=& \det(U_{\rm BdG}){\rm Pf}[{\cal D}(k_x) \tau^x] \nonumber\\
&=& (-1)^n~{\rm Pf}[{\cal D}(k_x) \tau^x]~.
\end{eqnarray}
The product inside the Pfaffian reads 
\begin{eqnarray}
{\cal D}(k_x) \tau^x &=& 
\begin{pmatrix}
E(k_x)   &  i\Delta~U^\dag \sigma^y U^* \\
-i\Delta~U^T \sigma^y U & -E(k_x)
\end{pmatrix}
\begin{pmatrix}
0 & \openone\\
\openone & 0
\end{pmatrix}
\nonumber\\
&=&
\begin{pmatrix}
  i\Delta~U^\dag \sigma^y U^* & E(k_x)   \\
-E(k_x)                         &-i\Delta~U^T \sigma^y U 
\end{pmatrix}~,
\end{eqnarray}
where $E(k_x)$ is a diagonal matrix with entries given by $\{\varepsilon_\lambda +\lambda V + \varepsilon_0-\mu\mp2 t\}$, $\lambda=\pm1$. This matrix is indeed antisymmetric. One can easily see that, upon a reordering of the rows and columns described by a real unitary matrix ${\cal V}$, the matrix ${\cal D}(k_x) \tau^x$ can be put into a block diagonal form, where each block is a $4\times4$ matrix involving the particle states with eigenvalues differing only by the sign of $V$ and their hole counterparts. Namely, each block has the form
\begin{equation}
\begin{pmatrix}
0 	& -\Delta & \varepsilon + V  & 0 \\
\Delta & 0  & 0 & \varepsilon - V   \\
- \varepsilon - V  & 0  & 0 & \Delta \\
 0  & -\varepsilon + V & -\Delta& 0 \\
\end{pmatrix}~,
\end{equation}
where $\varepsilon =  \varepsilon_\lambda + \varepsilon_0-\mu\mp2 t$. Thus, this means that 
\begin{align}
&{\rm Pf}[{\cal H}_{\rm BdG}(k_x)\sigma^y\tau^y] = (-1)^n~{\rm Pf}[{\cal D}(k_x) \tau^x]\\
&= (-1)^n~\det({\cal V}) ~\prod_{\lambda}[V^2 - \Delta^2-(\mu - \varepsilon_\lambda - \varepsilon_0\pm 2 t)^2]~,\nonumber
\end{align}
where the upper and lower signs refer to $k_x=0$ and $k_x=\pi/a$, respectively and the matrix ${\cal V}$ is the same for both $k_x=0,\pi/a$.  Thus, we finally have
\begin{eqnarray}
{\cal Q} &=& {\rm sign}\left\{{\rm Pf}[{\cal H}_{\rm BdG}(0)\sigma^y\tau^y]~{\rm Pf}[{\cal H}_{\rm BdG}(\pi/a)\sigma^y\tau^y]\right\}\nonumber\\
&=& \prod_{\lambda,\eta=\pm1}{\rm sign}\left[\Delta^2 + (\mu - \varepsilon_\lambda - \varepsilon_0+\eta~2 t)^2 -V^2\right]\nonumber\\
\end{eqnarray}
and consequently the phase boundaries are given by Eq.~\eqref{eq:phasebound}. We finally mention that analogous results have been found in Ref.~\onlinecite{ZhouPRB2011} in the case of spinless fermions in a $p$-wave superconducting nanowire.

\section{Transport across a NS junction}\label{app:B}

%
\begin{figure}
\includegraphics[width=1.0\linewidth]{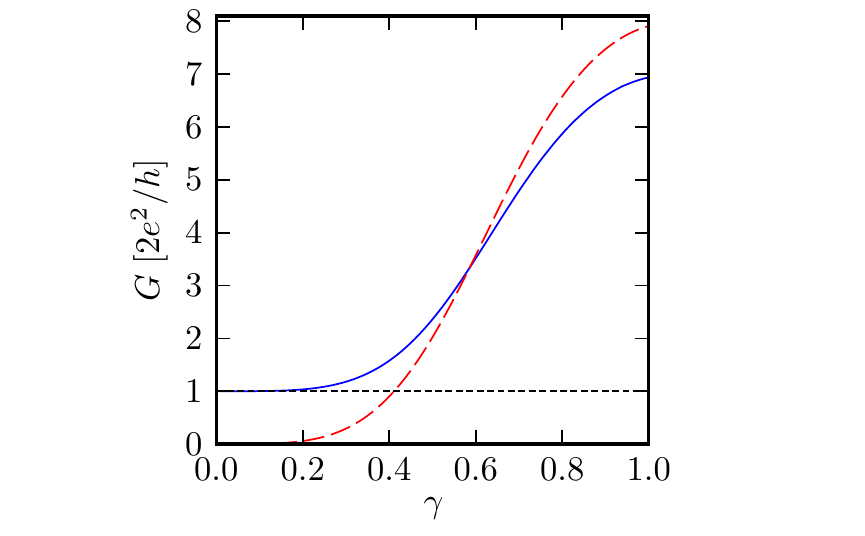}
\caption{(Color online) Conductance (in units of $2e^2/h$) of the NS junction as a function of the parameter $\gamma$ controlling the transparency of the barrier. The (blue) solid line refers to a S-nanowire in the topologically non-trivial phase ($V/t=0.2$ and $\mu/t=1.1$), while the (red) dashed line to a S-nanowire in the topologically trivial phase ($V/t=0.2$ and $\mu/t=1.5$).
\label{fig:transport}}
\end{figure}

In this Appendix we investigate another important tool to assess the topological phase of an S-nanowire coupled to a normal electrode: the low-bias conductance across a NS junction in the tunneling limit. In the limit of low bias, transmission through the superconductor is completely suppressed and the conductance of the NS junction can be expressed in terms of the Andreev reflection matrix $r_{he}$ (at the chemical potential) 
\begin{equation}
G = \frac{2e^2}{h}~{\rm Tr}\left[r^{\dag}_{he}r_{he}\right]=\frac{2e^2}{h}~\sum_{m=1}^{N_{\rm oc}}R_m~.
\end{equation}
Here $R_m$ are the eigenvalues of the Hermitian matrix $r^{\dag}_{he}r_{he}$ and $N_{\rm oc}$ is the number of open channels in the normal electrode. Owing to particle-hole symmetry, the $R_m$'s are either two-fold degenerate or equal to $0$ or $1$~\cite{BeriPRB2009,WimmerNJP2011}. The presence of a fully Andreev-reflected mode (giving a quantized contribution to the conductance) is a signature of the existence of an uncoupled Majorana fermion at the Fermi energy~\cite{FlensbergPRB2010, LawPRL2009}. As a consequence it is possible to write the conductance in the following form~\cite{WimmerNJP2011}
\begin{equation}
G = \frac{2e^2}{h} \left(1-{\cal Q} + 4{\sum_m}' R_m\right)~,
\end{equation}
where the primed sum is restricted to the degenerate Andreev reflection eigenvalues and ${\cal Q}$ is the topological invariant in Eq.~\eqref{eq:topinvariant}. In the limit of poorly transparent barriers ($\gamma\ll1$), we expect that almost all modes are fully reflected ($R_{m}\approx0$, though never exactly zero~\cite{BeriPRB2009,WimmerNJP2011}) and thus 
\begin{equation}\label{eq:Gtunneling}
G\approx \frac{2e^2}{h}\left(1-{\cal Q}\right)~.
\end{equation}
As a consequence the low-bias conductance of the NS junction in the tunneling limit gives an important information on the topological phase of the S-nanowire~\cite{SenguptaPRB2011,LawPRL2009,FlensbergPRB2010}. This result can be extended to almost transparent barriers if we include a ballistic quantum point contact close to the NS interface~\cite{WimmerNJP2011}. In Fig.~\ref{fig:transport} we show the conductance $G$  (in units of $2e^2/h$) as  a function the parameter $\gamma$ controlling the transparency of the barrier. In the tunneling limit ($\gamma\ll1$), we notice that the conductance approaches $0$ for a topologically trivial S-nanowire (red dashed line, $V/t=0.2$ and $\mu/t=1.5$) or $2e^2/h$ for a topologically non-trivial S-nanowire (blue solid line, $V/t=0.2$ and $\mu/t=1.1$), in agreement with Eq.~\eqref{eq:Gtunneling}.



\begin{thebibliography}{77}
\bibitem{note:gamma}
	A simple relation between the transmissivity ${\cal T}$ of the barrier and the parameter $\gamma$ can not  be in general derived. We simply notice that in the tunneling limit ($\gamma\ll1$) we expect that ${\cal T} \propto \gamma^2$.
\bibitem{note:pwave}
	To be more precise, the superconducting pairing induced in a one-dimensional wire with SO coupling is purely of the type $(p_x + i p_y)$ only if the chirality component of the dispersion curve with higher energy can be neglected. This happens when $V\gg\Delta$. Moreover, an increase of the Zeeman field produces a decrease of the effective gap $E_{\rm g}$, reinforcing the validity of the Andreev approximation.
\end{thebibliography}
\end{document}